\documentclass[fleqn,twoside]{article}
\usepackage{espcrc2}
\usepackage{fix2col}

\usepackage{graphicx}
\usepackage[figuresright]{rotating}

\newcommand{\AmS}{{\protect\the\textfont2
  A\kern-.1667em\lower.5ex\hbox{M}\kern-.125emS}}

\def\epem{\ensuremath{\mathrm{e}^+\mathrm{e}^-}}
\def\ra{\ensuremath{\rightarrow}}
\def\Zo{\ensuremath{\mathrm {Z}}}
\def\Ho{\ensuremath{\mathrm {H}}}
\def\rts {\ensuremath{\sqrt{s}}}
\def\MZ{\ensuremath{m_{\mathrm{Z}}}}%
\def\clb{\ensuremath{\mathrm{CL}_\mathrm{b}}}
\def\clsb{\ensuremath{\mathrm{CL}_\mathrm{s+b}}}
\def\cls{\ensuremath{\mathrm{CL}_\mathrm{s}}}
\def\MH{\ensuremath{m_{\mathrm{H}}}}%
\def\pb{\mbox{pb$^{-1}$}}

\def\GeV{\ifmmode {\mathrm{\ Ge\kern -0.1em V}}\else
                   \textrm{Ge\kern -0.1em V}\fi}%

\newlength{\figwidth}
\usepackage{amsmath}
\title{Search for the Standard Model Higgs Boson at LEP}

\author{Andr\'e G. Holzner\address{Institut f\"ur Teilchenphysik, 
    ETH H\"onggerberg\\
    CH-8093 Z\"urich, Switzerland\\
    Email: {\tt Andre.Holzner@cern.ch}\\
    }}
       
\begin{document}

\begin{abstract}
One of the missions of the LEP program was the search for the Standard
Model Higgs Boson. The skillful operation of the machine in the year 2000,
the final year of operation, has allowed the four collaborations
ALEPH, DELPHI, L3 and OPAL to collect 536 \pb{} of data at
center-of-mass energies of 206 \GeV{} or higher. 
This data is used to probe the existence of
the Higgs boson up to a mass of around 115 \GeV. 
Tantalizing candidates have been observed in excess over
the Standard Model predictions, but without enough statistical power to
claim a discovery. A Higgs boson lighter than 114.4 \GeV{} is hence excluded
at 95\% confidence level.

\vspace{1pc}
\end{abstract}

\maketitle

\section{Introduction}

The Standard Model (SM) of electroweak interactions 
is one of the most successful theories: It describes the physics
at the highest energy scales known today with unprecedented
precision. However, the SU(2) local gauge symmetry breaking (or
equivalently, the existence of massive fermions and gauge bosons) is
still not understood. The {\sl Higgs mechanism}~\cite{higgs} is the most widely
accepted mechanism allowing for massive fundamental particles while
preserving the local gauge invariance of the theory. It predicts a new
scalar particle, the {\sl Higgs boson}. Its mass is essentially
unknown, theoretical considerations give an upper limit of about
800 \GeV{}~\cite{Hambye:1997ax}. Recent fits of electroweak
parameters to the data favor a light Higgs ($\MH < 193 \GeV$ at 95\%
confidence level)~\cite{higgs-ew-limit}.

\section{Experimental Signatures and Event Selection}

\begin{figure*}[htb]
\vspace{9pt}
\begin{center}
  \includegraphics[width=0.9\figwidth]{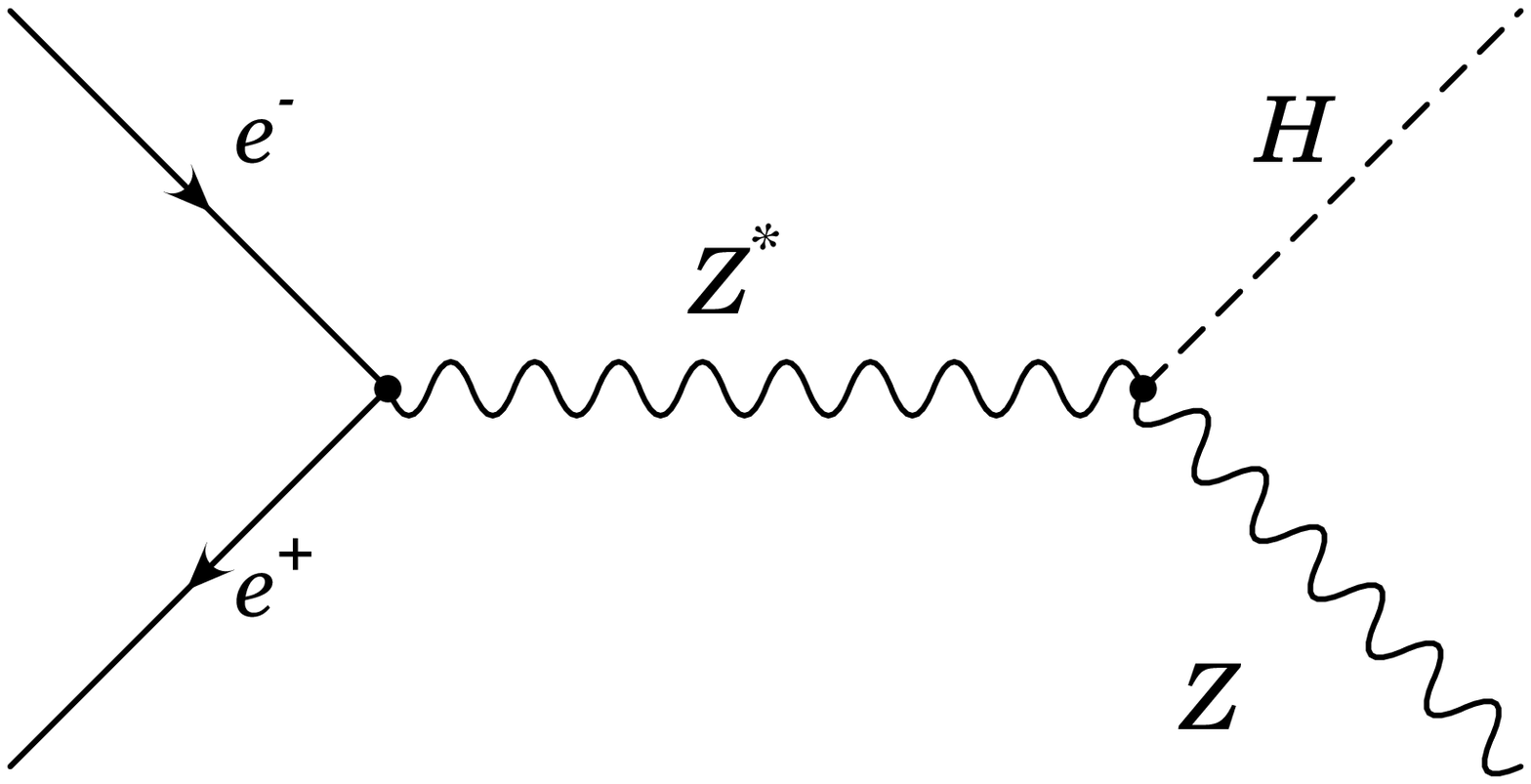}\qquad
  \includegraphics[width=0.9\figwidth]{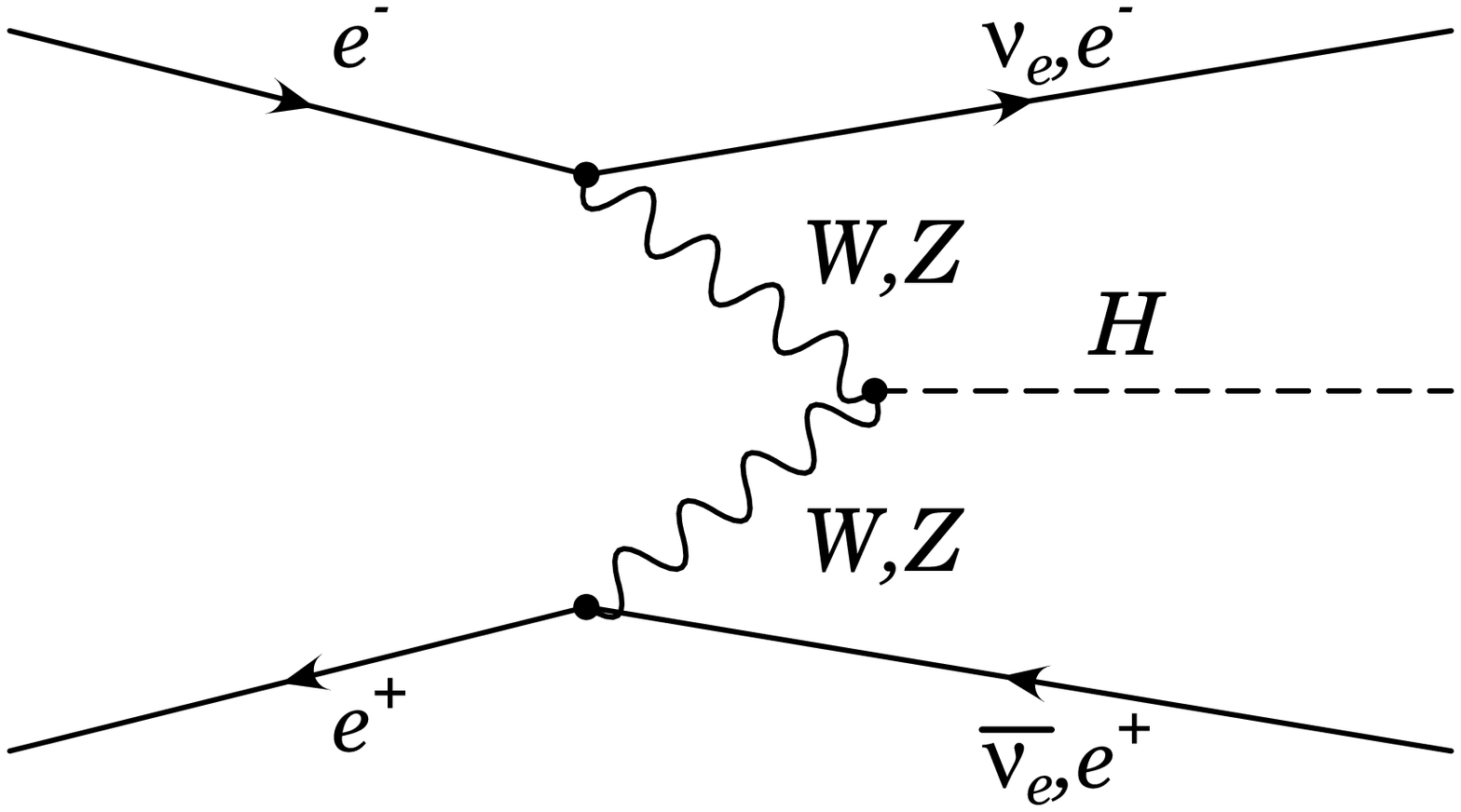}
\end{center}
\caption{The dominant Higgs production mechanisms in \epem{}
  collisions at LEP energies: Higgs strahlung (left) and weak boson
  fusion (right).
}
\label{fig:higgs-prod}
\end{figure*}

\begin{figure*}[h!]
\vspace{9pt}
\begin{center}
  \includegraphics[width=1.1\figwidth]{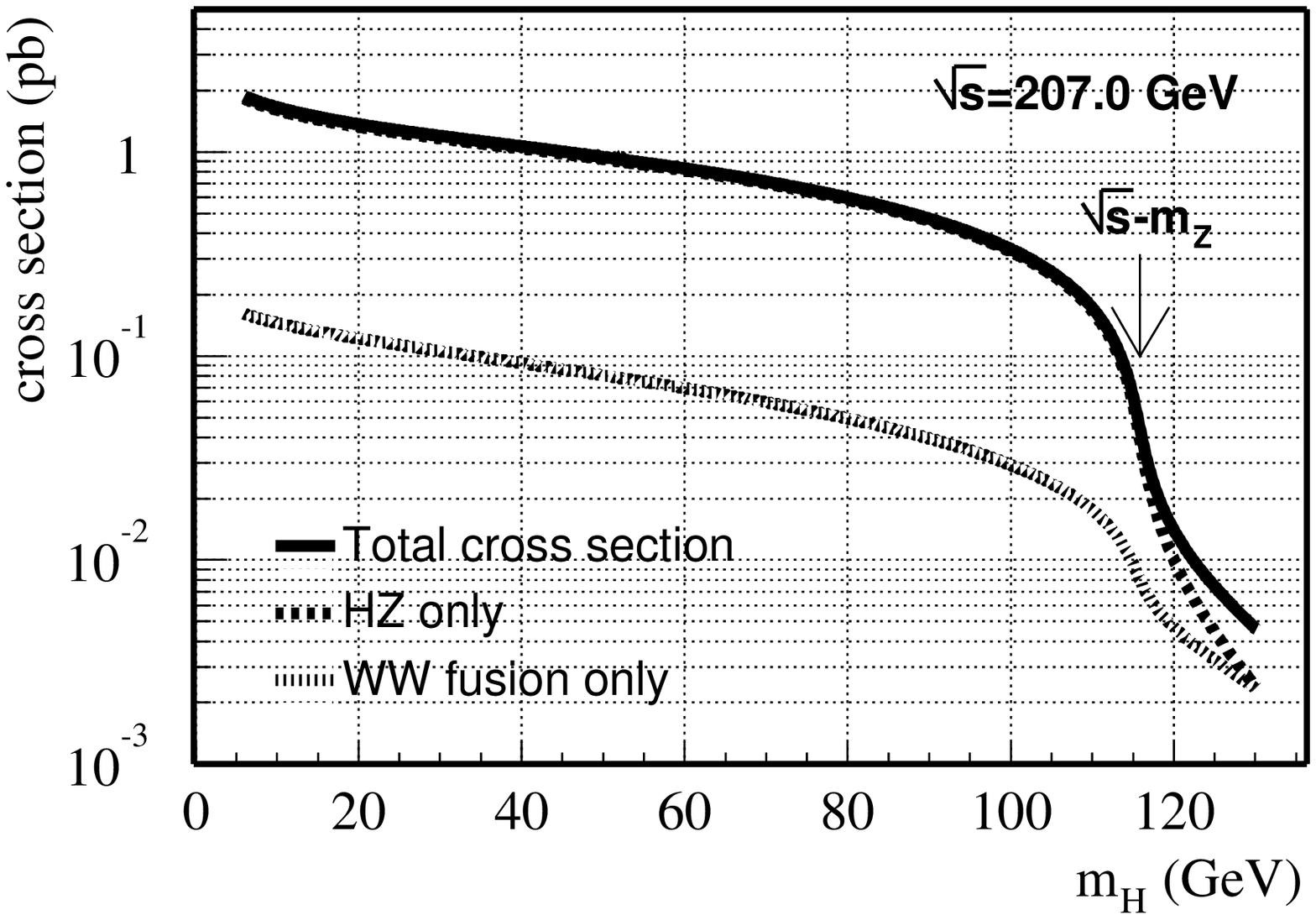}
  \includegraphics[width=1.1\figwidth]{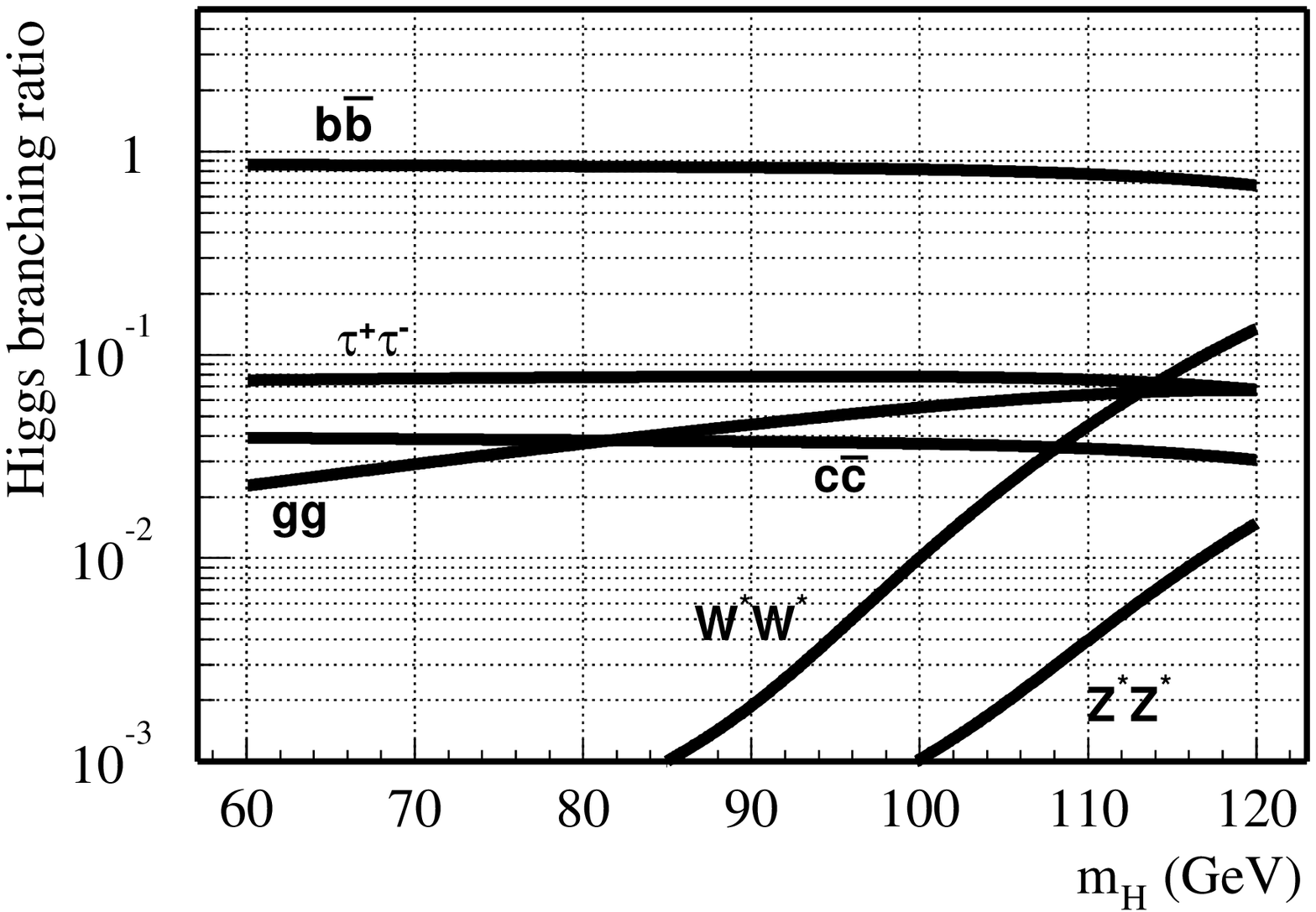}
\end{center}
\caption{Higgs production cross section at LEP for a center-of-mass 
energy of 207 \GeV{} as function of the Higgs mass (left). The solid line 
shows the total cross section, the fine dashed line the cross section
of the W fusion diagram and the coarse dashed line the
cross section from the Higgs strahlung process. 
The arrow indicates the {\sl kinematic limit} $\rts - \MH$.
The right plot shows the Higgs branching ratios as function
of the Higgs mass. For masses reachable at LEP, the
$\mathrm{b\bar{b}}$ decay mode is dominant, while $\tau^+\tau^-$ is
the second most abundant.
The cross sections and branching ratios were calculated using the
HZHA generator~\cite{hzha}.
}
\label{fig:higgs-xsect-br}
\end{figure*}

The dominant production mechanisms at LEP are the {\sl Higgs
  strahlung} $\epem \ra \Ho\Zo$ and the {\sl weak boson fusion}
(Fig.~\ref{fig:higgs-prod}). The Higgs strahlung diagram is the
dominant one up to Higgs masses of about $\rts - \MZ$ where the fusion
diagram becomes important. The cross section as function of the Higgs
mass for a center-of-mass energy of 207 \GeV{} is shown in
Fig.~\ref{fig:higgs-xsect-br} (left).

The Higgs boson decays mainly into a
pair of the heaviest particle allowed by kinematics. 
This is due to the fact that the
couplings of particles to the Higgs are the stronger the higher their
mass. Fig.~\ref{fig:higgs-xsect-br} (right) shows the branching
ratios as function of the Higgs mass. In the mass range accessible at
LEP, the Higgs decays mainly into a pair of b-quarks, e.g.\  74\% at
$\MH = 115 \GeV$.

The search analyses are inspired by the \Ho\Zo \ra $\mathrm{b\bar{b}} \Zo$
final state, although the fusion diagram and the interference with the
Higgs strahlung diagram and non-b decay modes are included in the
efficiency calculation. The experimental signatures are determined by
the Higgs and \Zo{} decay modes. 
In the following paragraphs, the main search channels are summarized.

\subsection{Four-jet channel}

The high branching ratio $\Zo \ra \mathrm{q}\bar{\mathrm{q}}$ of about
70\%  makes this channel the most abundant one. The most important
backgrounds are W pair production ($\sigma \sim 17 \mathrm{pb}$) where one di-jet pair
happens to have a mass close to the \Zo{} mass and $\epem \ra
\mathrm{q}\bar{\mathrm{q}}$ ($\sigma \sim 81 \mathrm{pb}$) with 
hard gluon radiation.

\subsection{Jets + missing energy channel}

Events in this channel are characterized by a missing mass close to the
\Zo{} mass and two acoplanar b-jets. The major backgrounds are 
$\epem \ra \mathrm{q}\bar{\mathrm{q}}\mathrm{e}\nu$ ($\sigma \sim 3.2
\mathrm{pb}$) where one electron escapes 
detection along the beam pipe and 
$\epem \ra \mathrm{W}^+\mathrm{W}^- \ra \mathrm{q}\bar{\mathrm{q}}\mu\nu$ or
$\epem \ra \mathrm{W}^+\mathrm{W}^- \ra \mathrm{q}\bar{\mathrm{q}}\tau\nu$ 
events where the lepton is not
identified. Pair production of Z bosons ($\sigma \sim 1.3
\mathrm{pb}$) and the process 
$\epem \ra \mathrm{q}\bar{\mathrm{q}}(\gamma)$ with undetected
initial state radiation of photons constitute further important backgrounds.

\subsection{Jets + leptons channel}

Events in this channel have two leptons with a dilepton invariant mass
close to the \Zo{} mass and two b-jets. The relatively low branching ratio of 
$\Zo \ra \ell^+\ell^-$, 3.4\% per lepton flavor, makes this a
small statistics channel. However, the most important background is
\Zo{} pair production (where one \Zo{} decays into quarks and the other
decays into leptons) which also has a small cross section, thus making this
signature extremely clean. All four experiments also include the case
where the Higgs decays into a pair of $\tau$ leptons and the \Zo{}
decays hadronically.

\begin{figure}[h!]
\vspace{9pt}
\begin{center}
  \includegraphics[width=\figwidth]{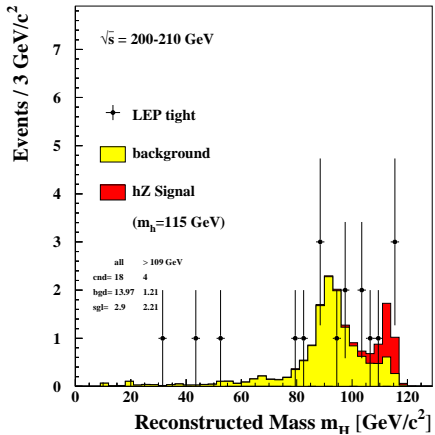}
\end{center}
\caption{Higgs candidate mass ($\MH^\mathrm{rec}$) distribution after
  tight cuts for the combination of all four LEP experiments and all
  search channels~\cite{lephiggs}. The light
  histogram represents the expected background, the dark histogram is
  the signal (\MH = 115 \GeV) 
  on top of the background while the dots are the data. The cuts were
  chosen such that the ratio of the number of expected signal and background events
  above $\MH^\mathrm{rec}$ = 109 \GeV{} is approximately 2:1 (2.21
  signal and 1.21 background events where 4 events are observed in data).
}
\label{fig:lep-mass}
\end{figure}

Fig.~\ref{fig:lep-mass} shows the reconstructed Higgs mass of all LEP
experiments and all search channels combined at a level of the
selection where the number of expected 
signal events above 109 \GeV{} is about twice the number of expected
background events in the same region.
Note that these mass distributions are {\sl
not} used directly to draw quantitative conclusions about the presence or
absence of a signal. This is done by combining the reconstructed mass 
with other event properties.

\section{Statistical Method and Quantitative Results}

\begin{figure*}[htb]
\vspace{9pt}
\begin{center}
  \includegraphics[width=\figwidth]{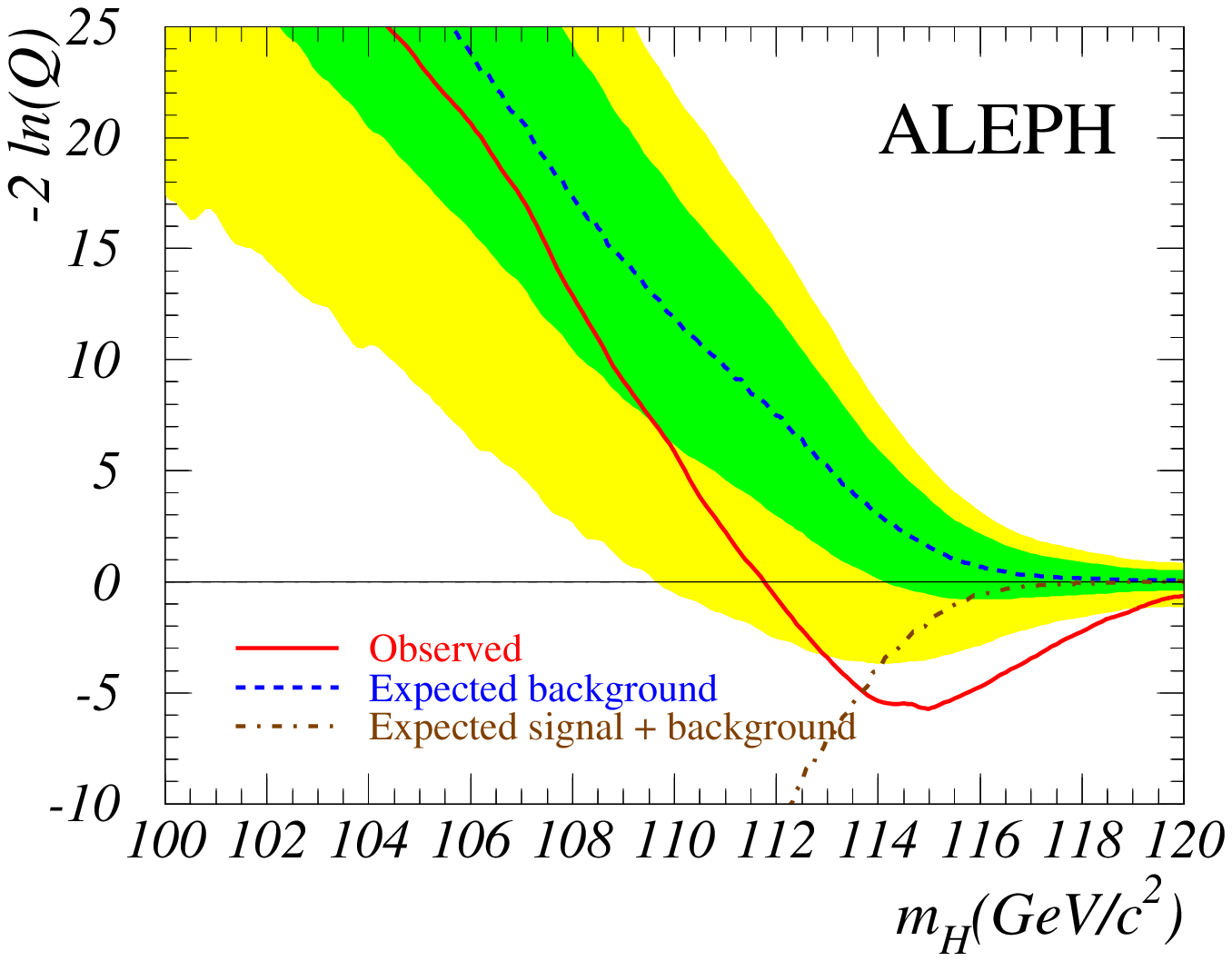}
  \includegraphics[width=\figwidth]{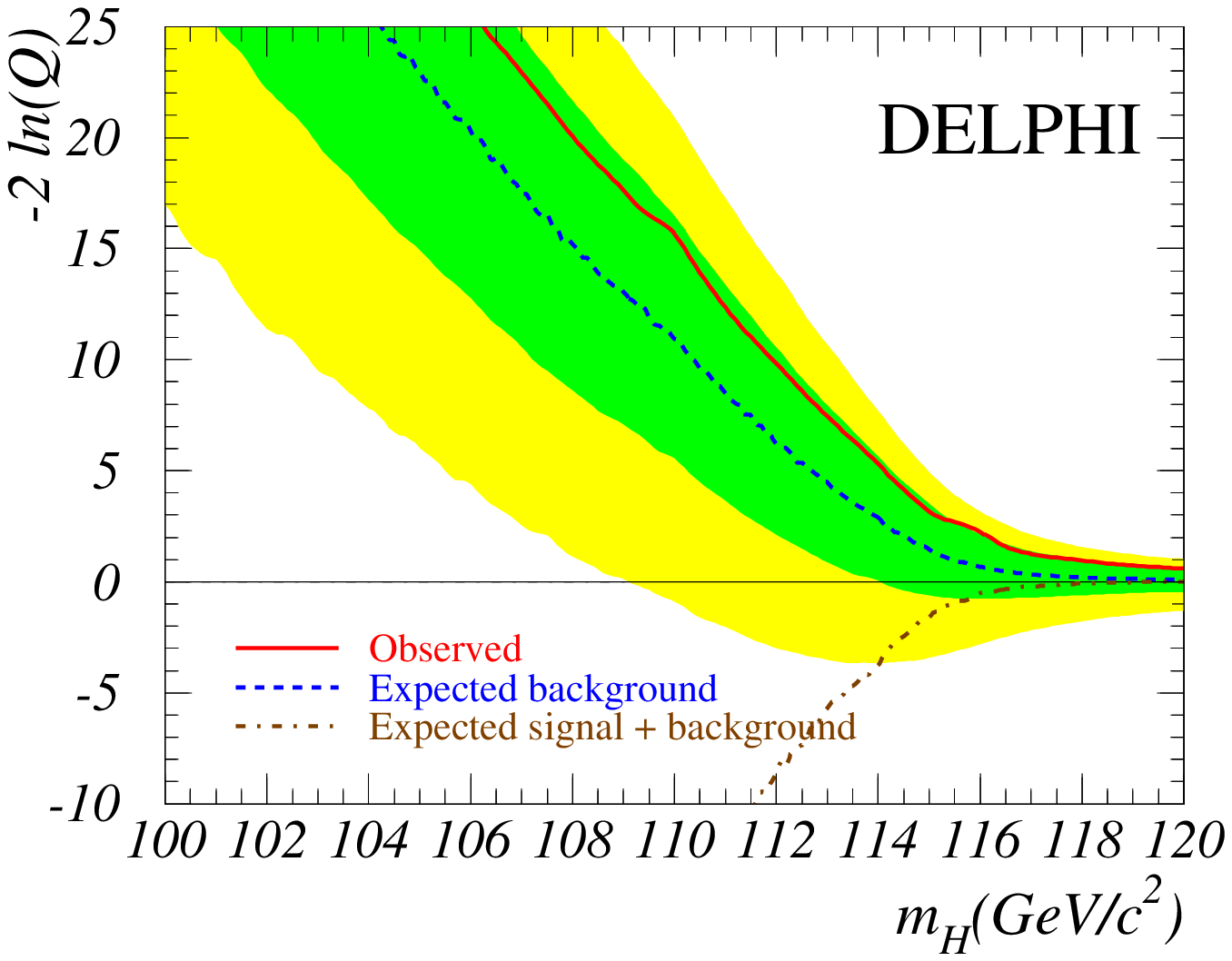}\\
  \includegraphics[width=\figwidth]{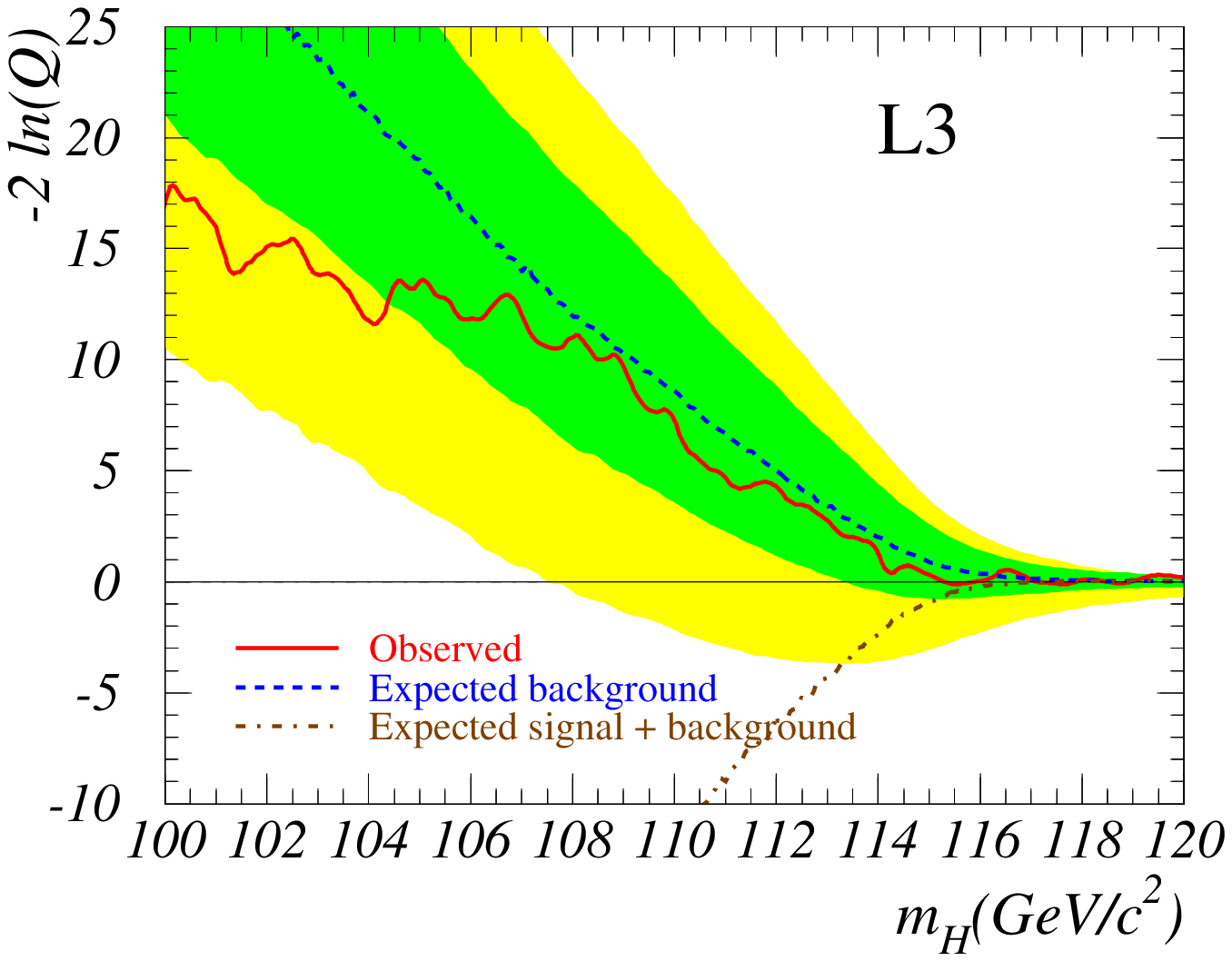}
  \includegraphics[width=\figwidth]{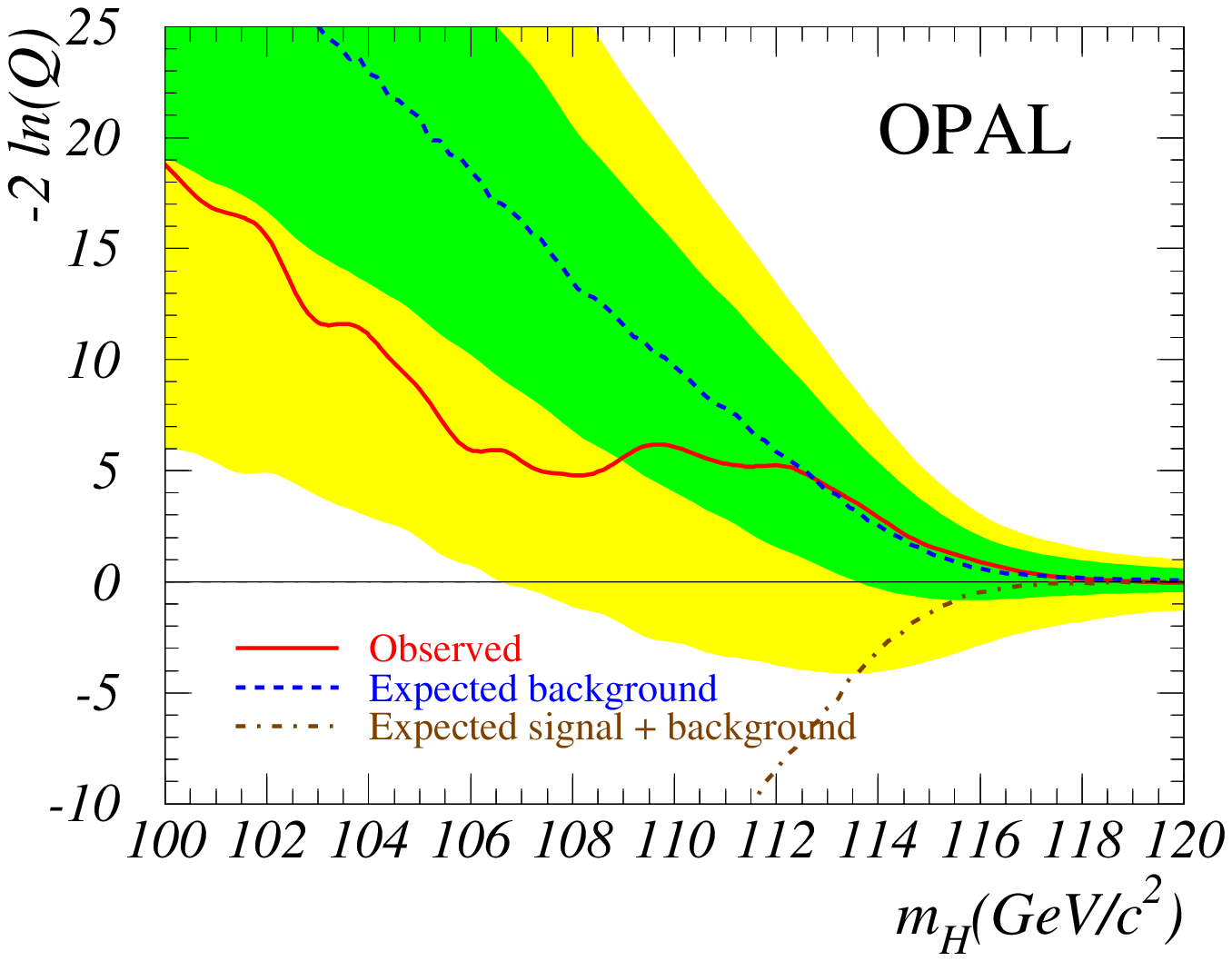}
\end{center}
\caption{Evolution of $-2 \ln Q$ as function of the
  Higgs mass hypothesis of the individual LEP experiments for all
  search channels combined~\cite{lephiggs}. 
The solid line corresponds to the observed data,
the dashed line to the expected background and the dash-dotted line
to the expected signal+background (for a sliding Higgs mass
hypothesis). The dark and light bands indicate the 1 and 2$\sigma$
deviations from the background. 
}
\label{fig:experiments-lnq}
\end{figure*}

\begin{figure}[h!]
\vspace{9pt}
\begin{center}
  \includegraphics[width=\figwidth]{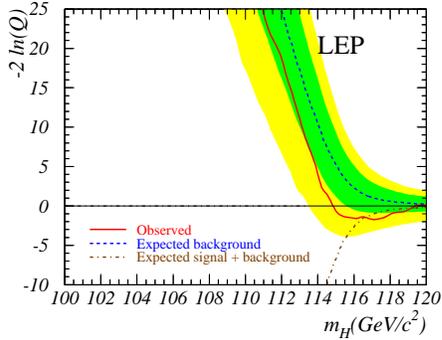}
\end{center}
\caption{Evolution of $-2 \ln Q$ as function of the
  Higgs mass hypothesis of all four LEP experiments and all search
  channels combined~\cite{lephiggs}. The meaning of the lines and bands is the same as
  in Fig.~\ref{fig:experiments-lnq}.
  The data is never more than
  two standard deviations away from the background over the mass range
  shown. 
  The distance between the background and signal+background lines is a
  measure of the sensitivity (or separation power between background
  and signal+background) of the analysis.
  The observation crosses the expectation from signal+background
  around 116 \GeV, however the separation power is very small at this mass.
}
\label{fig:lep-lnq}
\end{figure}

After a more or less tight event selection, usually a {\sl final discriminant} is constructed
from several variables of which each shows some separation power
between signal and background. The distribution of this final
discriminant is used to calculate the following likelihood ratio $Q$:

\begin{equation}
  \label{eq:llr}
  Q(x) = \dfrac{{\cal L}(x|s+b)}{{\cal L}(x|b)}
\end{equation}
where ${\cal L}(x|s+b)$ is the likelihood of $x$ under the
signal+background hypothesis (similarly for ${\cal L}(x|b)$). If the
final variable distribution is binned into a histogram, ${\cal L}$ is the
product of the corresponding Poisson probabilities over all histogram bins.
The variable $x$ usually represents the observed data but $Q$ can also be
calculated for the expected background.
Fig.~\ref{fig:experiments-lnq} shows the evolution of 
$-2 \ln Q$ as function of the Higgs mass hypothesis
for the individual LEP experiments.
The ALEPH result shows some excess, mainly from the four-jet channel,
in the mass region around 115 \GeV, which is however not observed by the
other experiments. 
Fig.~\ref{fig:lep-lnq} shows the combination of all LEP experiments
and search channels. The data are on the signal-like side of the
background but are never more than $2\sigma$ away from it.

Confidence levels are calculated by comparing the data to a large
number of Gedanken experiments, obtained from the expected signal and
background final discriminant distributions: \clb{} is the fraction
of {\sl background} experiments being more background like ($Q < Q_\mathrm{data}$) while \clsb{} is the fraction of
{\sl signal+background} experiments being more background like than
the data. Values of $1-\clb < 5.7 \cdot 10^{-7} (2.7 \cdot 10^{-3})$
are interpreted as $5\sigma$ discovery ($3\sigma$ evidence). 
Higgs masses for which $\cls = \clsb / \clb < 0.05$ are excluded at
95\% confidence level.

\begin{figure}[h!]
\vspace{9pt}
\begin{center}
  \includegraphics[width=\figwidth]{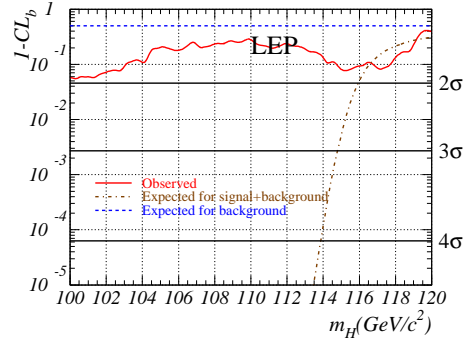}
\end{center}
\caption{Probability $1-\clb$ as function of the Higgs mass
  hypothesis. The solid line corresponds to the observed data, the
  dash-dotted line corresponds to the expectation from a signal (at
  each mass hypothesis) and the dashed horizontal line at 
  $1-\clb = 0.5$ is the expectation from background. The solid
  horizontal lines indicate the probabilities for a 2,3 and 4$\sigma$
  deviation from the background.}
\label{fig:lep-clb}
\end{figure}

The evolution of $1-\clb$ as function of the Higgs mass hypothesis is
shown in Fig.~\ref{fig:lep-clb}. The values of $1-\clb$ at \MH = 116
\GeV{} for the combination and the individual experiments are given in
Table~\ref{tab:clb}.  

\begin{table}[htb]
\begin{center}
  \begin{tabular}{lr@{.}l}
                  & \multicolumn{2}{c}{$1-\clb$} \\ \hline
 ALEPH            & 0&00241 \\
 DELPHI           & 0&874 \\
 L3               & 0&348 \\
 OPAL             & 0&543 \\ \hline
 Four-jet         & 0&0570 \\ 
 All but four-jet & 0&368 \\  \hline
 LEP              & 0&099  \\
  \end{tabular}
\end{center}
\caption{Probability $1-\clb$ at \MH = 116 \GeV{} for the individual
  experiments, the LEP combined result and the four-jet and all other
  channels~\cite{lephiggs}. Small values indicate a deviation from the
  background in the signal-like direction.
}\label{tab:clb}
\end{table}

\section{Summary}

A search for the Standard Model Higgs Boson at LEP
using 2461 \pb{} of data (at center-of-mass energies of 189 \GeV{} and
higher) was performed. In the highest energy data sample, the ALEPH
collaboration observed some highly significant candidates around a
Higgs mass of 115 \GeV{} in the four-jet channel. The probability for
the background to generate such a fluctuation is 0.24\%. However, the
other LEP experiments do not observe any candidates with a local signal
over background ratio greater than one. 
A lower bound on the Higgs mass is set at
114.4 \GeV{} at 95\% confidence level~\cite{lephiggs}.

All the results quoted here are preliminary, however since 
two of the four experiments published their final
analyses~\cite{aleph_final,l3_final}, one has submitted its final
publication~\cite{opal_final} and one submitted its final results to
conferences~\cite{delphi_ichep2002}, the final combined LEP result is expected to
be similar.


\end{document}